\documentclass[12pt,twoside]{article}
\usepackage{fleqn,espcrc1}
\usepackage{latexsym} 
\usepackage{amssymb}  
\usepackage{amsfonts} 
\usepackage{multirow}
\usepackage{epsf}
\usepackage{graphicx}
\def\y0{y^{(0)}}

\newcommand \beq{\begin{eqnarray}}
\newcommand \eeq{\end{eqnarray}}

\newcommand{\mnote}[1]{\marginpar{\tiny {}}}   

\def \jpsi {\mbox{J/$\Psi$}}

\def \pt   {\mbox{$p_{\rm t}$}}

\begin{document}

\title{\bf Physics of Ultra-Relativistic Nuclear Collisions with Heavy Beams at LHC Energy
}
\author{
   Peter Braun-Munzinger\\
   Gesellschaft f{\"u}r Schwerionenforschung\\
   64220 Darmstadt, Germany\\
   }
\maketitle

\begin{abstract}

    \noindent 
    We discuss current plans for experiments with ultra-relativistic
    nuclear collisions with heavy beams at LHC energy ($\sqrt{s} =
    5.5$ TeV/nucleon pair). Emphasis will be placed on processes which are
    unique to the LHC program. They include event-by-event
    interferometry, complete spectroscopy of  
    the $\Upsilon$ resonances, and open charm and open beauty measurements.

\end{abstract}


\section{Introduction and Physics Scenario}

\begin{figure}[thb]

\vspace{-1cm}

\epsfxsize=11cm
\begin{center}
\hspace*{0in}
\epsffile{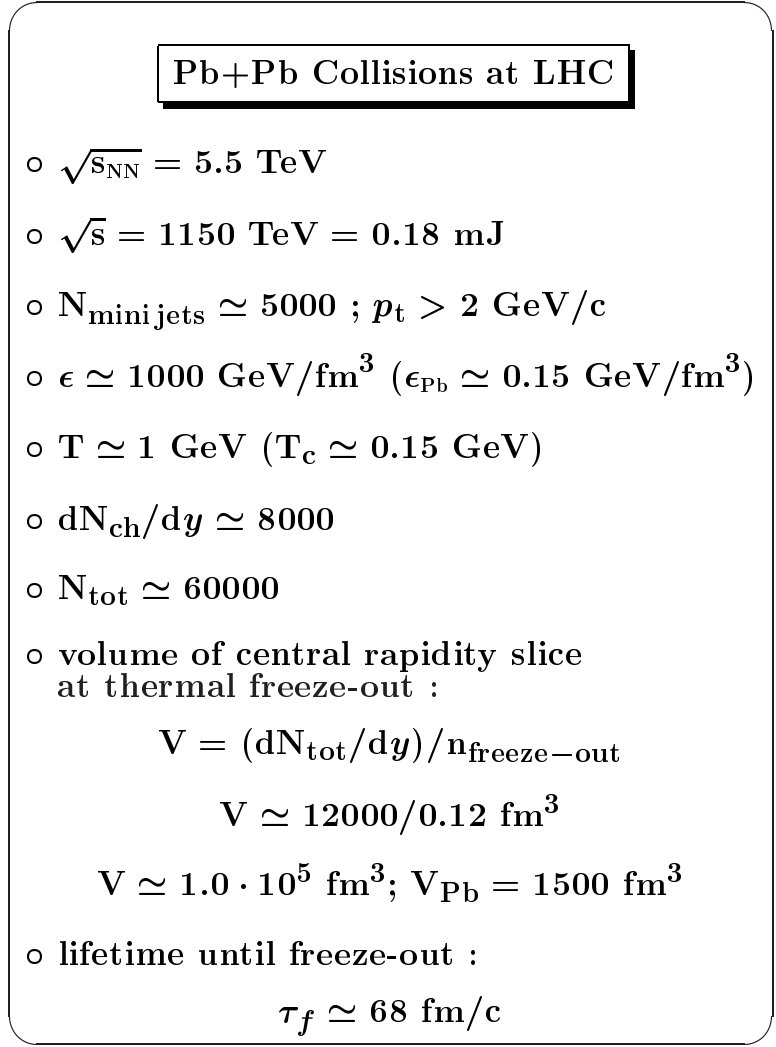}
\end{center}

\vspace{-1.5cm}

\caption{
Some typical parameters for Pb-Pb collisions at LHC energy
} 
\label{fig:phys_1}
\end{figure}

\noindent At LHC energy hard parton-parton collisions will provide the
dominant part of the transverse energy produced in a Pb-Pb collision
\cite{eskola}. Therefore, the initial stage of such a collision will
be manifestly partonic. Because of the very large cross section for
gluon-gluon scattering the gluons will reach equilibrium quickly and
initial temperatures of about 1 GeV at timescales $\ll 1$ fm/c will be
reached. Typical parameters of such partonic fireballs are collected
in Fig. ~\ref{fig:phys_1}.  The temporal evolution of such partonic
fireballs is depicted in more detail in Fig. ~\ref{fig:phys_2}. For
simplicity a first order phase transition was assumed to take place at
T$_c = 170$ MeV but the general picture should be independent of
this. In the calculation I have assumed a Bjorken-type longitudinal
expansion coupled with a transverse expansion which is small in the
parton and mixed phase but significant in the hadronic phase. Entropy
is conserved throughout the expansion.  Transverse expansion
parameters are given in the figure.  Because of the very large entropy
of dS/dy $\approx 41000$ \footnote{The corresponding quantity for
central Pb+Pb collisions at SPS energy is about 1800.}created in the
collision the life time of the fireball until thermal freeze-out is
about 68 fm/c, i.e. comparable to typical low energy nuclear physics
time scales \footnote{This is roughly equal to the oscillation time of
the giant dipole resonance in a Pb-nucleus.}.

Inspection of these figures shows that quark-gluon plasma will indeed
be produced over very large space-time volumes at the LHC. Under this
scenario we expect exciting and qualitatively new physics results from
ultra-relativistic nuclear collisions at the LHC even 6 years after
the first collisions at RHIC have been measured. At the LHC there will
be two experiments where heavy ion collisions can be studied. The
ALICE experiment \cite{alice1,alice2,alice3} is dedicated to the
study of heavy ion collisions and designed such that both hadronic and
leptonic observables can be measured. Pb+Pb collisions can also be
studied in the CMS experiment \cite{cms}, albeit with focus on high
mass di-lepton spectroscopy. Rather than describe these experiments in
detail I will concentrate in the following on selected physics topics
which are unique and particular to experiments at LHC energy.  Other
possible and interesting investigations are described in
\cite{alice1,alice2,alice3}.

\section{Hadronic Observables}
\noindent
A number of interesting physics topics can be 
investigated by looking at the produced hadrons. They include:
\begin{itemize}
\item spectra and production yields of multi-strange baryons.
\item event-by-event fluctuations in the hadronic final state.
\item azimuthal anisotropies and elliptic flow.
\item jet physics via particle spectra at high transverse momentum.
\end{itemize}

\begin{figure}[thb]

\vspace{-1.5cm}

\epsfxsize=14cm
\begin{center}
\hspace*{0in}
\epsffile{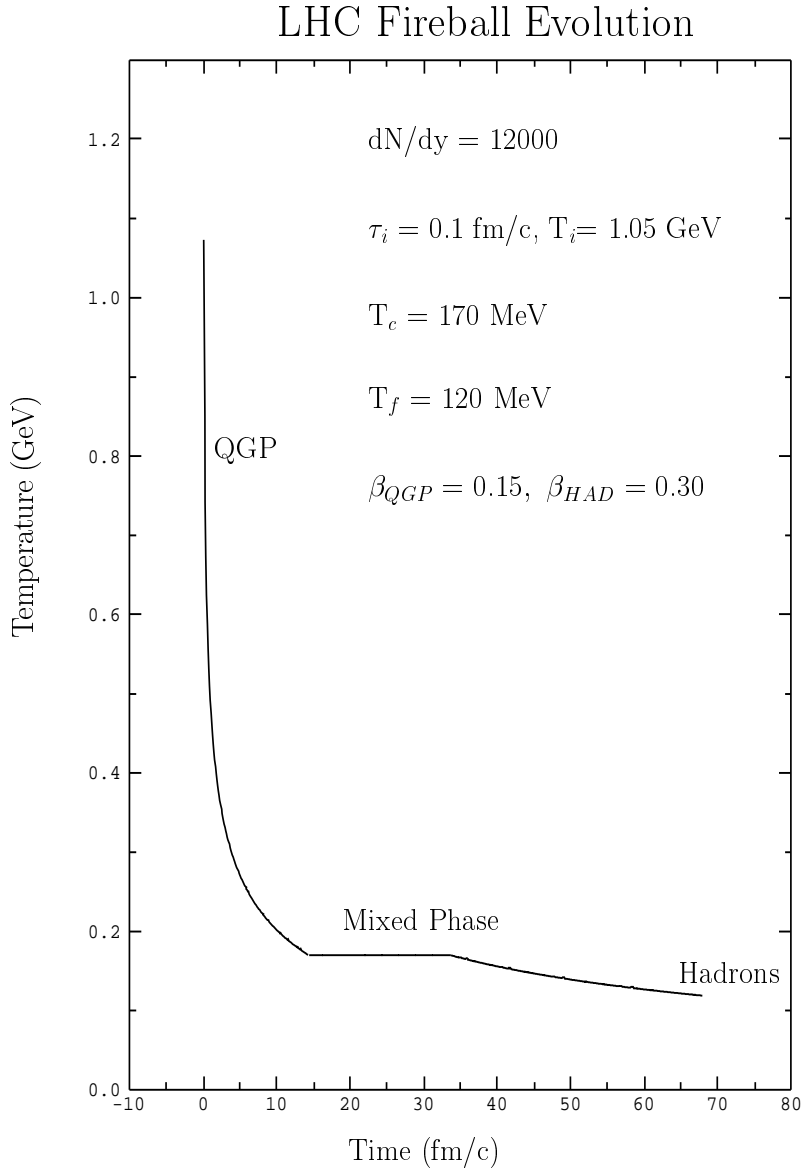}
\end{center}

\vspace{-4.8cm}

\caption{
Temporal evolution of the partonic fireball in a Bjorken scenario with
additional transverse expansion. 
}

\label{fig:phys_2}
\end{figure}

These hadronic measurements are of particular interest at LHC energies
since the above mentioned large entropy implies very large rapidity
densities for hadrons. Assuming that chemical freeze-out takes place
near the phase boundary as has recently been established for hadron
yields at AGS energy \cite{therm1} and SPS energy \cite{therm2,therm3}
I can use the same thermal model to predict rapidity densities for
Pb+Pb collisions at LHC energy. For the thermal model parameters I use
a temperature T=170 MeV (close to chemical freeze-out at SPS energies,
\cite{therm3}) and a baryon chemical potential $\mu_b$ = 10
MeV. Setting $\mu_b$ to zero changes the particle densities very
little. To get absolute rapidity densities one has, of course, to
specify the freeze-out volume. I take this from the calculation shown
in Fig. ~\ref{fig:phys_2}, which implies a fireball volume (for a one
unit of rapidity wide slice at mid-rapidity) at chemical freeze-out of
14400 fm$^3$. A very similar calculation has recently been worked out
for RHIC energy \cite{therm4}.

In Table \ref{tab:yields} we show the corresponding rapidity densities
for a number of particle species. The resulting numbers are
astonishing indeed. In particular, the high strangeness content of the
fireball leads to very large rapidity densities for strange and
multi-strange baryons. The sum of all strange baryons approaches 300
per event in one unit of rapidity, and 2000 per event overall! We note
that, while the particle ratios are a firm prediction within the
thermal model, absolute yields depend on the initial temperature of
the (partonic) fireball. Summing up the rapidity density of all
charged hadrons (after weak decays) leads then to dN$_{\rm ch}$/dy
=7560 [T$_i$(GeV)/1.05]$^3$ since the initial temperature in the present
calculation is close to 1.05 GeV (see Fig. ~\ref{fig:phys_2}).

\begin{table}[htb]
\newlength{\digitwidth} \settowidth{\digitwidth}{\rm 0}
\catcode`?=\active \def?{\kern\digitwidth}
\caption{Rapidity densities at mid-rapidity for Pb+Pb central 
collisions as predicted by thermal model calculations \cite{therm3}. 
Chemical freeze-out takes place at T=170 MeV in a volume of 14400 fm$^3$.
The  baryon chemical potential is assumed to be $\mu_b = 10$ MeV. 
For more details see text.
}
\label{tab:yields}
\begin{center}
\begin{tabular}{||c|c||}
\hline
&\\
particle species &dN/dy for T=170 MeV \\
 & $\mu_b$=10 MeV \\
\hline
\(\rm \pi^- \approx \pi^+ \) &  2500\\
\(\rm \pi^o \) &  2800 \\
\(\rm \eta \) & 270 \\
\(\rm \omega \)  &  220  \\
\(\rm \phi \)  &  57   \\
\(\rm K^+ \approx K^- \approx K^0_s \)  & 385  \\
\hline
\(\rm p \)  &  250  \\
\(\rm n \)  &  240 \\
\(\rm \overline{p} \)  &  220 \\
\(\rm \overline{n} \)  &  210 \\
\(\rm p-\overline{p} \)  &  30 \\
\(\rm \Lambda \)  & 126 \\
\(\rm \overline{\Lambda} \)  & 116 \\
\(\rm \Lambda(1405) \)  &  7 \\
\(\rm \Xi^- \approx \Xi^+ \)  & 17 \\ 
\(\rm \Omega^- \approx \Omega^+ \)  & 3 \\
\(\rm d \)  &  1.0 \\
\(\rm \overline{d} \)  &  0.9 \\
\hline
\end{tabular}
\end{center}
\end{table}

With such hadron yields\footnote{The coverage of the ALICE central
  detector is $\pm 0.88$ units of pseudo-rapidity centered at
  90$^{\circ}$ to the beams.} one can perform a number of exciting
investigations, as demonstrated with the list above. For example,
Hanbury-Brown/Twiss interferometry measurements for pions and kaons
can be performed on an event-by-event basis. Within the above fireball
evolution scenario one expects (two-dimensional) transverse radius
parameters of the order of 22 fm, i.e. more than a factor of 2 larger
than currently measured at SPS energy. The corresponding width in
momentum space of the two-particle correlation function will then be
of the order of 10 MeV or less, a real challenge for experiments. One
should however keep in mind that the large radial flow velocities
expected for central Pb-Pb collisions will effectively reduce the
"apparent" radii, leading to wider correlation functions if the
transverse momentum of the particle pair is not too small. With such
measurements one can not only determine the temporal and spatial
evolution of the fireball but also search for fluctuations which are
expected should the phase transition be of second order. As a
"by-product" of such measurements one can also obtain information
about the final state interaction of the particles involved. In
particular, this may provide the unique possibility to determine the
interaction between strange baryons, as well as to search for
resonances in these systems. Both programs would be of high interest
for the hadron physics community.

\section{Quarkonia, Photonic  and Leptonic Observables}

\noindent

Many exciting and important investigations will become possible by studying
leptonic observables in Pb+Pb collisions at LHC energy. They include:
\begin{itemize}
\item complete quarkonium spectroscopy.
\item D- and B-meson spectroscopy via semi-leptonic decay channels.
\item study of $\omega$ and $\phi$-meson production.
\item measurement of the thermal lepton pair continuum at high masses.
\item search for anomalous enhancement in the lepton pair continuum at
  low masses.  
\item measurement of Z$_0$ production.
\end{itemize}

In addition, the PHOS detector within the ALICE experiment can be used
to measure direct photons with the aim to study:
\begin{itemize}
\item thermal emission from the quark-gluon plasma and from the mixed
  phase.
\item photon-photon correlations.
\end{itemize}

In the following I will discuss the new possibilities for quarkonium
spectroscopy and D- and B-meson spectroscopy. Another look at leptonic
observables within the ALICE forward muon spectrometer can be found in
\cite{morsch}.

\begin{figure}[htb]

\vspace{-0.5cm}

\epsfxsize=14cm
\begin{center}
  \epsffile{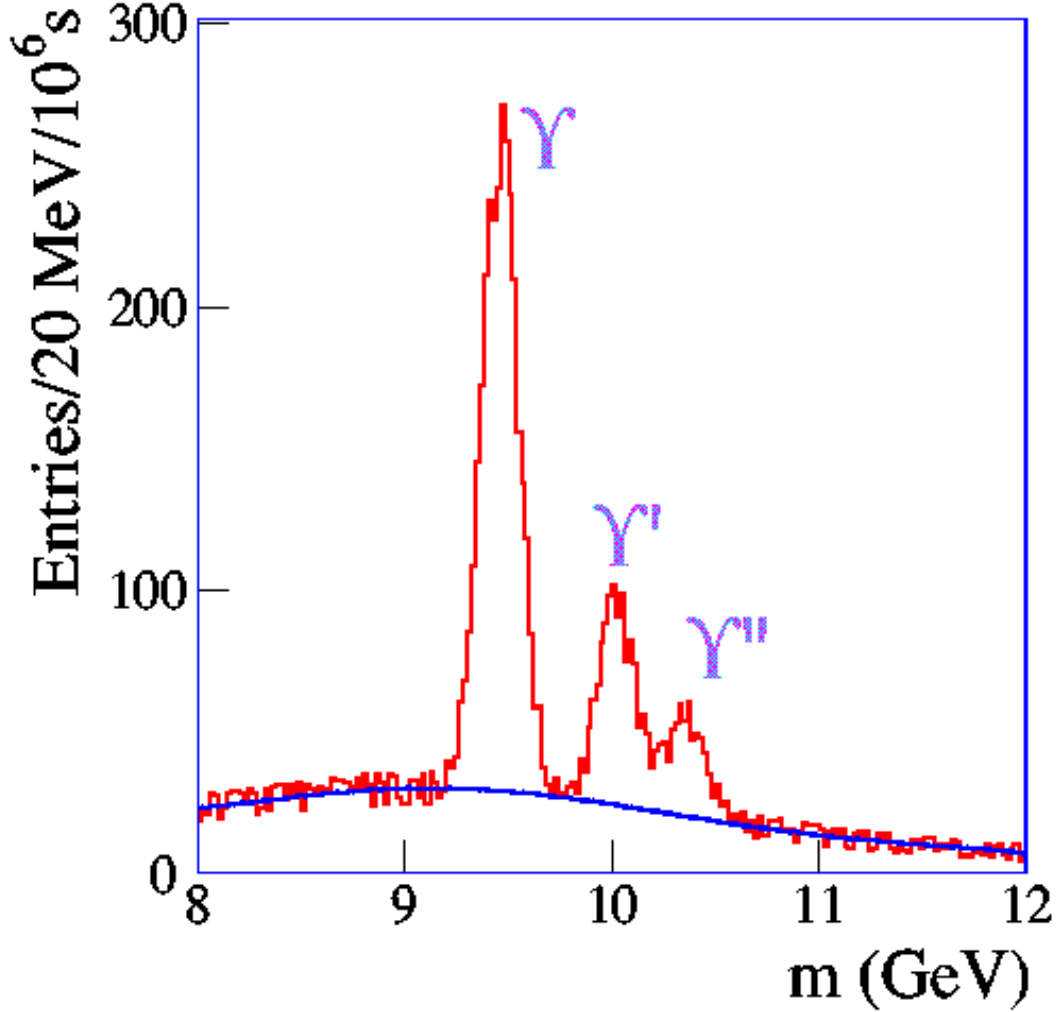}
\end{center}

\vspace{-2cm}

 \caption{
Expected dimuon mass spectrum for the ALICE muon arm after a running
time of 10$^6$ s \cite{alice2,morsch}. No plasma suppression is taken into account in this
simulation.} 
 \label{fig:upsilon}
\end{figure}

The long life time of the partonic phase as is visible from Fig.
~\ref{fig:phys_1} and the high energy density of the partonic
fireballs should lead to complete suppression through Debye screening
of the color interaction in the deconfined and dense plasma even of
tightly bound resonances such as the $\Upsilon$ mesons.  At the LHC
one should be able to perform a complete quarkonium spectroscopy
(J/$\Psi, \Psi', \Upsilon, \Upsilon', \Upsilon''$) to measure the
sequential melting of the resonances, on account of their different
radii, as a function of the energy density. Note that the $\Upsilon$,
because of its small radius of about 0.2 fm, should only "melt" at
energy densities above 30 GeV/fm$^3$ corresponding to temperatures
T$>$ 400 MeV and should therefore not be suppressed at RHIC
energies. The variation of the energy density can be achieved either
by changing the centrality (impact parameter) or the system size or
both. Particularly interesting is the case of the three $\Upsilon$
resonances whose radii vary between 0.2 and 0.8 fm (for comparison, the
radius of the J/$\Psi$ meson is about 0.45 fm). The particular
interest to study the suppression  for the
$\Upsilon$ family has been in detail discussed in \cite{gunion}.

\begin{figure}[htb]

\vspace{-0cm}

\epsfxsize=14cm
\begin{center}
  \epsffile{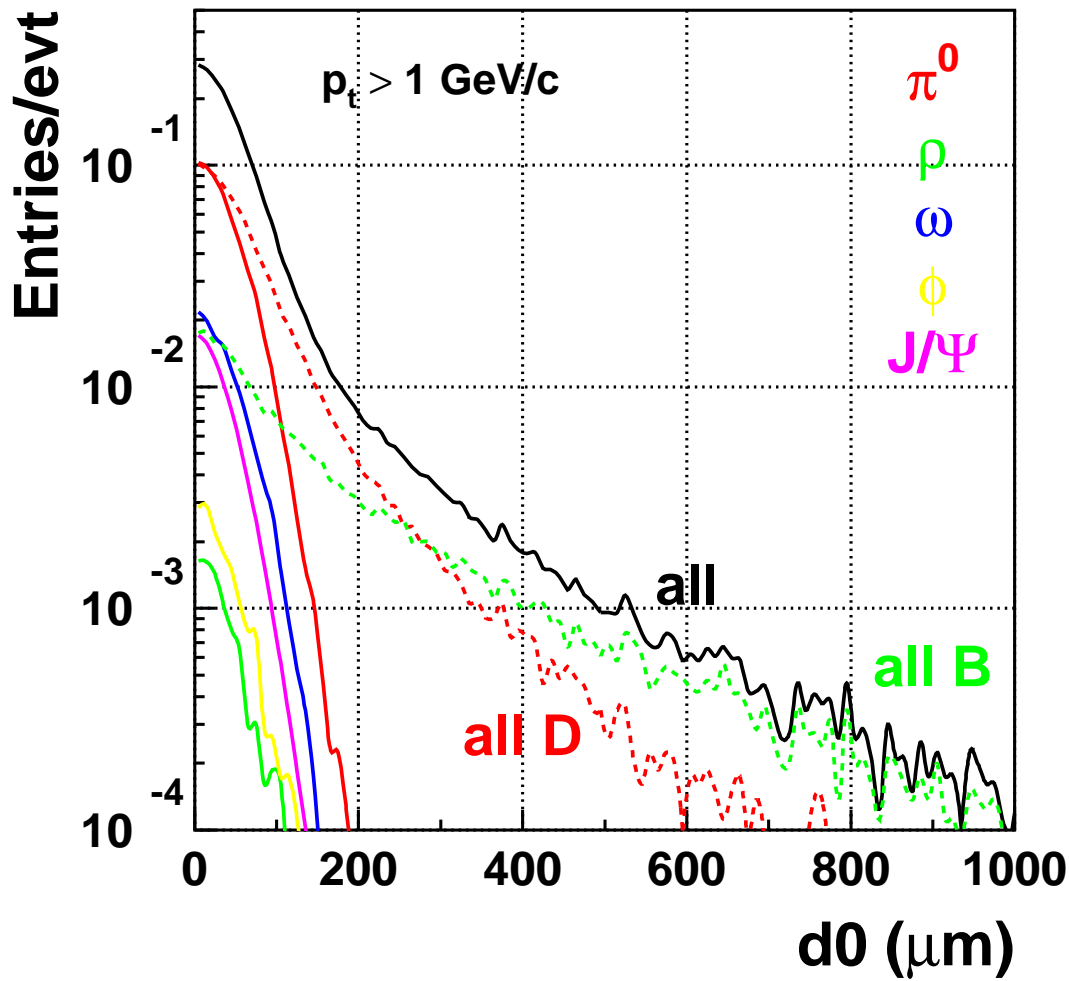}
\end{center}

\vspace{-2cm}

 \caption{
Distribution of the closest distance d$_0$ to the primary  vertex of
electron tracks for 
Pb+Pb collisions at the LHC. The long tails due to electrons from
semi-leptonic D- and B-decays are clearly visible. For details see text.}
 \label{fig:d0_all}
\end{figure}

At the LHC the $\Upsilon$ resonances can be measured in Pb+Pb
collisions in the CMS experiment \cite{cms}, in the ALICE muon arm
\cite{alice2,morsch} and in the ALICE central arm using the newly
added transition radiation detector \cite{alice3}. All three detectors
will provide 
a mass resolution of about 100 MeV or better for masses around 10 GeV/c$^2$,
sufficient for a good separation of the three resonances. A good
impression of the expected quality of the measurement is obtained by
looking at Fig.  \ref{fig:upsilon}. The continuum in this
figure is due to combinatorial background of muons from $\pi$ and K
decays, from semi-leptonic D-decays and, dominantly, from semi-leptonic
B-decays.

The large yields (in the absence of plasma suppression) for $\Upsilon$
states is of course connected to the expected copious production of
open charm and open beauty at LHC energies.  The measurement of the
cross sections for open charm and open beauty production is
consequently of paramount importance for understanding the role of
color screening in the expected suppression of quarkonium production
in Pb+Pb collisions at LHC energy. First, one should note that
quarkonium production and open charm or open beauty production are
intimately related \cite{hard_probe}. The heavy quarks are dominantly
produced in hard partonic scattering processes (gluon fusion).  Some
of these heavy quarks will eventually form quarkonia, while the large
majority will end up in correlated pairs of D and B mesons. Typical
numbers are about 50 ${\rm c\bar c}$ pairs per unit of rapidity
compared to 0.5 for \jpsi\ mesons in central Pb+Pb collisions. These
numbers have been obtained \cite{alice3} using a perturbative QCD 
approach based on the PYTHIA framework with a K-factor adjusted to
reproduce B meson production in ${\rm p \bar p}$ collisions at the
Tevatron and D meson production at lower energies.  Similar results
have been obtained by Gavin et al. \cite{gavin}.  The extrapolation to
nucleus-nucleus collisions is done as in \cite{charm_pbm}, i.e. by the
total number of nucleon-nucleon collisions.  This number is calculated
from the nuclear collision geometry at a given impact parameter
assuming a Woods-Saxon nuclear density distributions.

One should note that these predictions are still rather uncertain.
First, there are indications from the measurements of NA50 at CERN of
a possible enhanced open charm production (relative to PYTHIA
predictions) already in Pb+Pb collisions at SPS energy
\cite{na50_charm}. Secondly, the produced heavy quarks may suffer
significant energy losses in the hot and dense fireball formed at LHC
energies, leading to considerable kinematical rearrangement of the
final rapidity and \pt\ distributions and, consequently, also to
potentially drastic changes in the dilepton invariant mass spectrum
from semi-leptonic decays
\cite{shuryak_charm,vogt_charm}. 
Clearly, a direct measurement of
charm and beauty is mandatory for the interpretation of the
quarkonium data at LHC energies.

Furthermore, since the quark and the gluon structure functions are
likely to be different at LHC energies, one cannot use the Drell-Yan
continuum as a convenient normalization for the \jpsi\ measurements.
In addition, the Drell-Yan cross section is expected to be completely
masked by the open charm continuum.
However, a direct
measurement of the open charm yield simultaneously with the yield of
quarkonia will provide a natural normalization and a gauge against
which to quantify the expected suppression of quarkonia.

\begin{figure}[htb]

\vspace{-1cm}

\epsfxsize=15cm
\begin{center}
  \epsffile{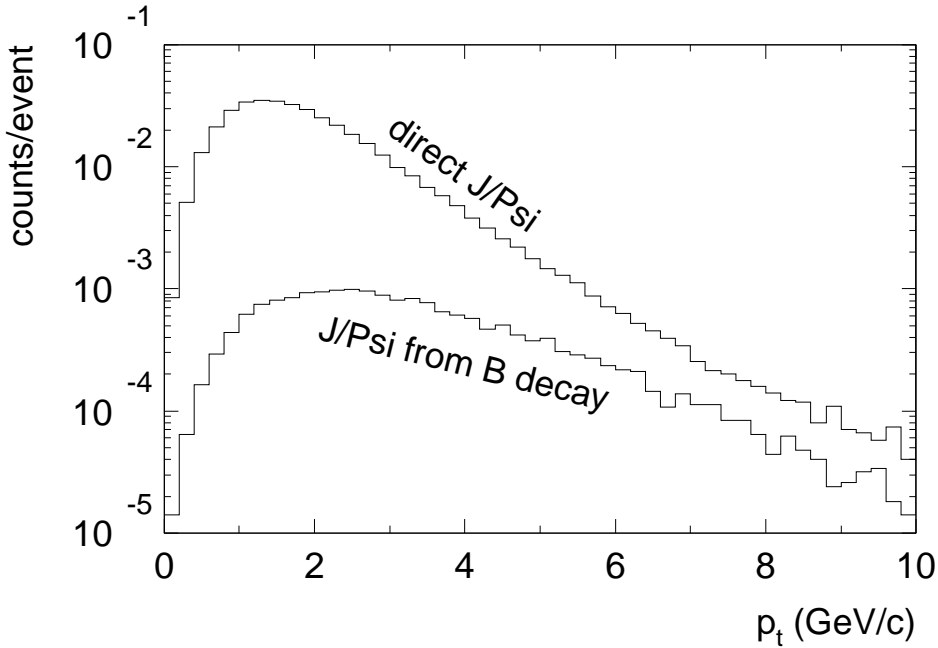}
\end{center}

\vspace{-2cm}

 \caption{
Comparison of yields for unsuppressed primary \jpsi\ production and
production via B-decay in Pb+Pb collisions at LHC energy. For more
details see text.}
 \label{fig:b-decay}
\end{figure}

Such a measurement can be obtained by making use of the transition
radiation detector in combination with the inner tracking systems and
time projection chamber of ALICE \cite{alice3}. The idea here is not to fully
reconstruct D- and B-mesons but to determine their yield and \pt\
spectra via the identification of displaced vertices of electrons and
positrons from semi-leptonic decays. Because of the life time on the
order of picoseconds the D- and B-mesons typically decay at a distance of a few
hundred $\mu$m away from the primary vertex. The quantity d$_0$, the
closest distance of a track to the true vertex, is plotted in
Fig. ~\ref{fig:d0_all} for Pb+Pb collisions. The performance of the
ALICE inner tracking system leads to a gaussian smearing of less than
100 $\mu$m for tracks with \pt $> 1$ GeV/c. This smearing is visible
for the prompt particles (electrons from Dalitz decays of neutral
mesons etc.). However, for distances above 200 $\mu$m the d$_0$
distribution is completely dominated by electrons from D- and
B-decay: with an appropriate cut one can isolate a nearly pure sample
of leptons from heavy meson decay. Many more details about these
procedures can be found in \cite{alice3}.

\begin{figure}[htb]

\vspace{-1cm}

\epsfxsize=14cm
\begin{center}
  \epsffile{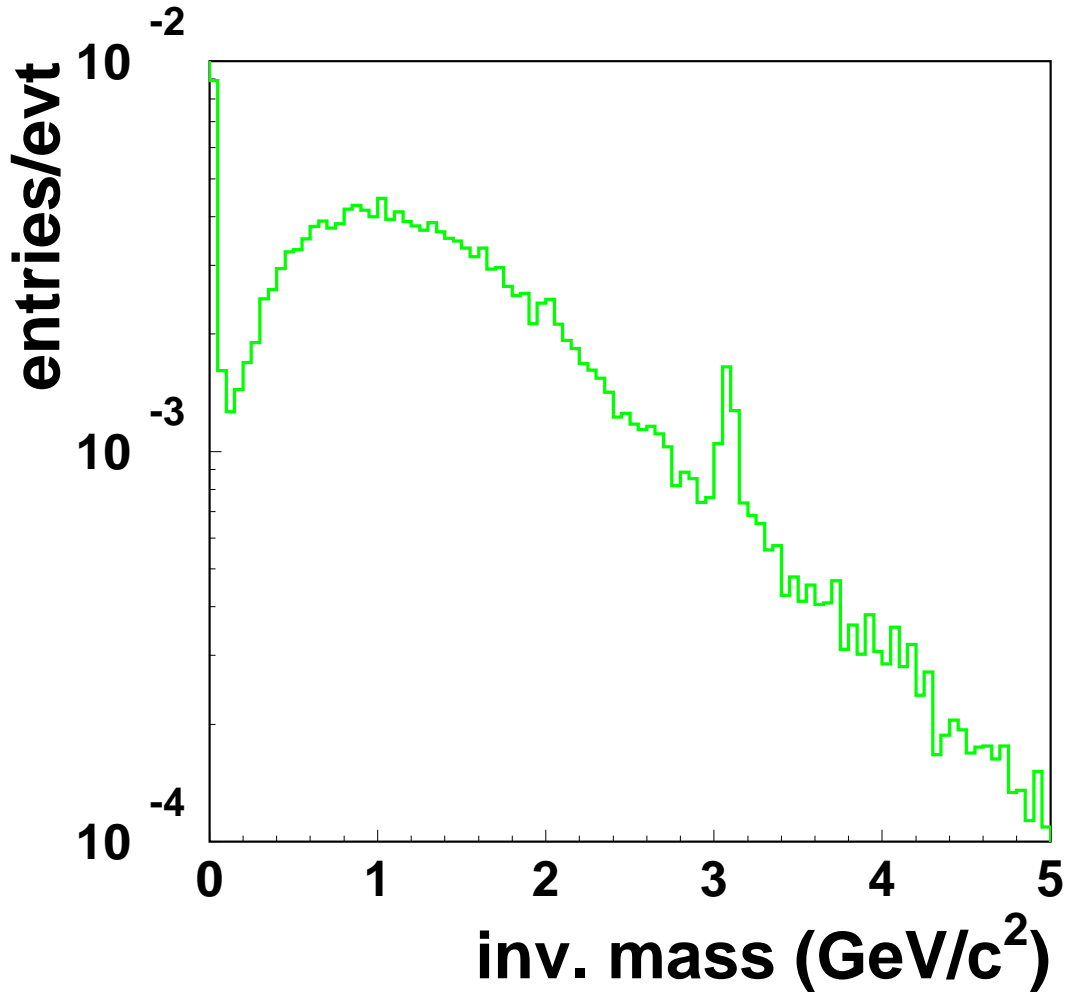}
\end{center}

\vspace{-1cm}

 \caption{
Reconstruction of \jpsi\ mesons from B-decay for Pb+Pb collisions in
the ALICE central detector. For details see text and \cite{alice3}.}
 \label{fig:secjpsi}
\end{figure}

Although the large open charm and open beauty cross sections make
these interesting measurements at least statistically straightforward,
they also lead to  further potential problems for the quarkonia
measurements. For example, neutral B-mesons have a 1.3\% branching ratio
to decay into \jpsi\ + X. The yields for such \jpsi\ mesons of course are not
plasma-suppressed and are not negligible compared to the
primary production of \jpsi\ mesons \footnote{A similar problem should also
  exist at RHIC energies and should be addressed there, too.}. This is
demonstrated in Fig. ~\ref{fig:b-decay}. Note that for these
calculations it was assumed that there is no plasma suppression for
the primary \jpsi\ mesons. From this figure it is clear that,
especially at high \pt\, secondary \jpsi\ production can obscure even
moderate plasma suppression of primary \jpsi\ mesons.

However, using similar techniques as for the D- and B-mesons one can
actually identify directly \cite{alice3} the secondary \jpsi\ mesons
in the ALICE central detector 
and separate 
them from primary production. Since the necessary vertex cut removes a
large amount of combinatorial background, the signal/noise ratio for
the identification of secondary \jpsi\ mesons is impressive, as
demonstrated in Fig. ~\ref{fig:secjpsi}. This technique not only
allows a clean separation of primary and secondary production of
\jpsi\ mesons but also provides a semi-exclusive measurement of the B$_0$
distribution.

Another issue to take into account at LHC energies is the secondary
production of \jpsi\ mesons from the 
annihilation of D mesons, i.e. the process D+$\bar {\rm D} \rightarrow
\jpsi + \pi$. Estimates for the yield due to this process have
recently been given \cite{ko_d,pbm_d} and were discussed at this
conference. For presently considered values of 
the cross section for D+$\bar {\rm D} \rightarrow \jpsi\ + \pi$
secondary production could possibly obscure the expected suppression in
the plasma. Especially for the $\Psi'$, secondary production could lead
to a rather dramatic enhancement\cite{pbm_d} with the $\Psi'$ yield
even exceeding the \jpsi\ yield. It is, therefore, clear
that a clean interpretation of \jpsi\ 
production data can only be obtained through a comprehensive measurement
of open charm production as is planned with the transition radiation
detector in the ALICE experiment.

\section{Summary and Outlook}
\noindent
The study of Pb+Pb collisions at LHC energy will provide qualitatively
new perspectives for ultra-relativistic heavy ion physics. As the
above discussions have shown, even 6 years after the begin of collider
experiments at RHIC there will be a rich and unique menue of
experiments to be performed on fireballs which are quasi-macroscopic
compared to normal nuclear parameters.

\section{Acknowledgements}

It is a pleasure to acknowledge many discussions with A. Andronic, P. Crochet,
A. Devismes, P. Gl\"assel, N. Herrmann, K. Redlich, and J. Stachel.

\end{document}